\documentclass[doublecol]{epl2} 

\renewcommand{\huge}{\large}
\renewcommand{\LARGE}{\large}
\renewcommand{\Large}{\large}
\usepackage{amsmath}
\usepackage{amssymb}
\usepackage{graphicx}
\usepackage{psfrag}
\usepackage[subnum]{cases}
\title{Number of adaptive steps to a local fitness peak}

\author{Kavita Jain}
\shortauthor{K. Jain}

\institute{                    
Theoretical Sciences Unit and Evolutionary and Organismal Biology
Unit, \\Jawaharlal Nehru Centre for Advanced Scientific Research,
Jakkur P.O., Bangalore 560064, India
}

\abstract{We consider a population of genotype
  sequences evolving on a rugged 
  fitness landscape with many local fitness peaks. The population
  walks uphill until it encounters a local fitness maximum. We find
  that the statistical properties of the walk length depend on whether
  the underlying fitness distribution has a finite mean. If the mean
  is finite, all the walk length cumulants 
  grow with the sequence length but approach a constant
  otherwise. Experimental 
  implications of our analytical  
  results are also discussed.}

\begin{document}

\def\be{\begin{equation}}
\def\ee{\end{equation}}
\def\bea{\begin{eqnarray}}
\def\eea{\end{eqnarray}}
\def\no{\nonumber}

\maketitle

\section{Introduction}

The  evolutionary process of adaptation is common in nature 
\cite{Orr:2005} and during the last decades, the dynamics of
adaptation have been studied in several  
experiments on microbial populations \cite{expt}. 
The nature of the adaptive process depends crucially on the
availability of beneficial mutations that improve the fitness
\cite{Jain1}.  If such mutations are
readily available as in populations of very large size, the
dynamics are well described by a deterministic 
theory \cite{qs} while for moderately large populations, a stochastic
theory which accounts for competing multiple mutations 
can be applied \cite{ci}. Here we work in the parameter 
regime where beneficial mutations are rare and 
a population of genotype sequences performs an 
{\it adaptive walk} on a fitness landscape \cite{Smith:1970,Gillespie:1991}. 

\begin{figure}
\begin{center}
\includegraphics[width=0.9 \linewidth,angle=0]{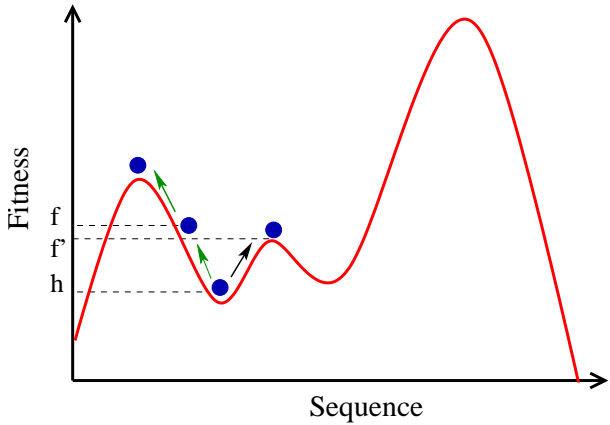}
\caption{(Color online) Schematic diagram to illustrate adaptive walk
  on a rugged fitness landscape with many local maxima. The population
  (filled circle) with fitness $h$ has fitter one-mutant neighbors with
  fitness $f, f', ...$. One of the better mutants is chosen with a
  transition probability
  (\ref{Tfh}). The global maximum is not accessible to the 
  population as it is not a 
  one-mutant neighbor and the walk terminates when the population reaches
  a local fitness maximum.}    
\label{adapwalk}
\end{center}
\end{figure}

More precisely, the adaptive walk model assumes that 
the number of mutants produced per generation is small so that
the population is genetically homogeneous and may be  
represented by a single particle. The weak mutation assumption also
renders the sequences differing by more than one mutation
inaccessible. Furthermore the sequences carrying mutations that
decrease the fitness do not survive and hence the adaptive walker
always walks uphill. On a rugged fitness landscape with many local
optima, the walk ends when a local fitness maximum is
encountered since a better fitness is at least two mutations
away as illustrated in Fig.~\ref{adapwalk}. Remarkably,
under these assumptions, the model depends only on a small set 
of parameters namely the sequence length and the 
fitness distribution underlying the fitness landscape. 
Recently some theoretical predictions for the first
step \cite{Orr} in the walk were tested in an experiment on a ssDNA
virus \cite{Rokyta:2005} and a reasonable agreement between theory and
experiment was found.  
As the adaptive walk describes a simple and biologically realistic
model of adaptation, it is important to analyse it 
in detail to extend our present understanding of adaptation dynamics. 

In this Letter, we focus on the statistical properties of the
length of adaptive walk defined as the
number of beneficial mutations accumulated until the population
reaches a local fitness maximum. Recently  the walk length distribution 
was calculated within an approximation for the model described
above \cite{Jain:2011d} and the mean walk 
length was computed exactly in a simplified
version of the adaptive walk \cite{Neidhart:2011}.   
However these studies assume that the
fitness distribution has a finite mean. Here we relax this
assumption and interestingly, we 
find that in the limit of infinitely long sequence, there is a 
transition in the behavior of the walk length distribution: it
vanishes for fitness distributions with finite mean but remains finite
otherwise. For finite sequences, this result implies that the 
walk length diverges with the sequence length for distributions with
finite mean. For such distributions, we show that all the walk length 
cumulants grow logarithmically with the sequence length and find the 
  proportionality constant for the first few cumulants. 
Our analytical results are compared with the
numerical results and their experimental implications are also
discussed.

\section{Model}

We work with binary sequences of length $L$ so that each sequence has $L$ 
neighbors which are one mutation away. As the fitness always increases
in an adaptive 
walk (see Fig.~\ref{adapwalk}), the mutants that lower the current
fitness $h$ of the walker are 
rejected and a mutant with given 
fitness $f > h$ is chosen with a transition probability $T(f, h|f)$
proportional to the fitness difference $f-h$ 
\cite{Gillespie:1991}. Thus the normalised transition probability is
given by 
\be
T(f, h|f)=\frac{f-h}{\sum_{g > h} g-h}
\label{Tfh}
\ee
where the fitnesses are independent random variables chosen from a
common distribution $p(f)$ with support on the interval $[0, u]$. 
Following previous works \cite{Joyce:2008,Neidhart:2011}, we choose
the fitnesses from a generalised Pareto distribution defined as
\be
p(f)= (1+\kappa f)^{-\frac{\kappa+1}{\kappa}}
\label{pf}
\ee
where the fitness $f$ is unbounded for $\kappa \geq
0$ and $f \leq -1/\kappa$ for $\kappa < 0$. The distribution of the
{\it beneficial} mutations is however governed by the upper tail of
the fitness distribution $p(f)$ \cite{Gillespie:1991} and hence can be
one of the three universal distributions only
\cite{Bouchaud:1990,Sornette:2000}. The fitness distribution $p(f)$
lies in the domain of  
the extreme value distribution given by Weibull distribution for
$\kappa < 0$, Gumbel distribution if  
$\kappa \to 0$ and   
  Fr{\'e}chet distribution for $\kappa > 0$. Although much of the
experimental data on distribution of beneficial mutations is
consistent with $\kappa \to 0$  
\cite{kappa0,Rokyta:2005}, recent works also support $\kappa < 0$
\cite{Rokyta:2008} and $\kappa >0$ \cite{Neidhart:2011}. 

The adaptive walk in the limits
$\kappa \to \pm \infty$ is well studied theoretically. When $\kappa
\to \infty$, 
the adaptive walk model reduces to a greedy walk \cite{Joyce:2008} for
which the walk length distribution is finite for infinitely long
sequences \cite{Orr:2003} while for 
$\kappa \to -\infty$, a random adaptive walk is obtained
\cite{Joyce:2008} for which the walk length distribution is a Poisson
distribution with mean $\ln L$ 
\cite{Flyvbjerg:1992}. Recently the adaptive walk model
  described above was studied in detail for $\kappa=-1$ and $\kappa
  \to 0$ and the walk length distribution was computed
  \cite{Jain:2011d}. Here we are 
interested in   
the properties of adaptive walk when $\kappa$ is arbitrary
  but finite.

Following \cite{Flyvbjerg:1992}, we consider the 
conditional probability ${\cal 
  P}_J(f)$ that the walker takes at least $J$ steps and has a fitness
$f$ at the $J$th step given that the initial fitness is
  $f_0$. For long sequences, one can write down the 
following recursion 
  relation for $J \geq 0$ \cite{Jain:2011d}:
\bea
{\cal P}_{J+1}(f,L)= \int_0^f dh ~L p(f) T(f, h|f) ~(1-q^L(h))
  {\cal P}_J(h,L) 
\label{PJfns}
\eea
where $q(h)=\int_0^h dg ~p(g)$. The above equation expresses the fact
that the walker can proceed to the next step if at least one fitness
value greater than the current fitness $h$ is available which occurs
with a probability $1-q^L(h)\approx 1-e^{- L \left(1+\kappa h
  \right)^{-\frac{1}{\kappa}}}$. The  walk length distribution $Q_J$ 
that {\it exactly} $J$ steps  are taken is related to ${\cal P}_J(f)$
according to the following relation \cite{Jain:2011d}:
\bea
Q_J(L) = \int_0^u dh ~q^L(h) {\cal
  P}_J(h,L)  
\label{QJdef} 
\eea
This is because in order to terminate the walk at the $J$th step, none
of the $L$ mutant fitnesses at the next step should exceed the fitness
at step $J$.  In the following, we set the initial fitness
  $f_0$ to be zero, ${\cal
  P}_0(f,L)=\delta(f)$ which ensures that the walker does 
not start at a local fitness maximum. 


\section{Transition in the behavior of walk length}

Using a scaling analysis and extreme value theory, 
we now show that the qualitative behavior of walk length 
distribution $Q_J$ changes  at 
$\kappa=1$. We find that the walk length
distribution vanishes for $\kappa < 1$ but remains finite for $\kappa
> 1$ as $L \to \infty$. We note that the behavior
of $Q_J$ discussed above  
for $\kappa \to \pm \infty$ is in accordance with our result. 

\begin{figure}
\begin{center}
\includegraphics[width=0.7 \linewidth,angle=270]{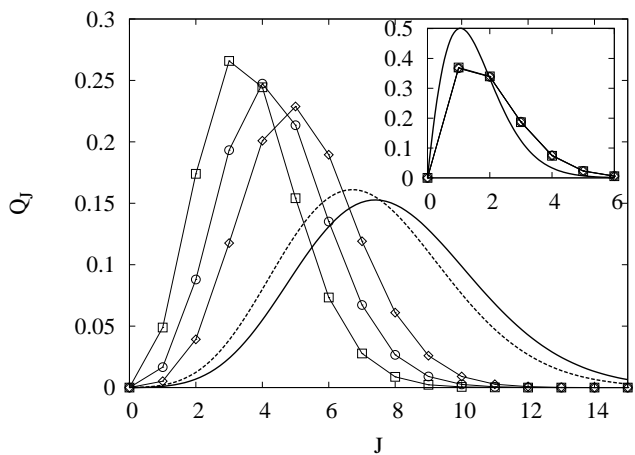}
\caption{Walk length distribution $Q_J$ as a function of $J$ for
  $\kappa=1/2$ (main) and $3/2$ 
  (inset) for $L=10^3$ (squares), $10^4$ (circles) and $10^5$
  (diamonds) to show that in the 
  infinite sequence length limit, $Q_J$ vanishes for $\kappa < 1$ but
  remains finite for $\kappa > 1$. The points  
  are joined to guide the eye. 
  For comparison, 
 the analytical result for the random adaptive walk (main) is shown for $L=500$ (broken line) and 
  $10^3$ (solid line) and for the 
  greedy walk (inset) for infinitely long sequence.}     
\label{behav}
\end{center}
\end{figure}

For $\kappa < 1$, it is a good approximation to replace the sum on the
right hand side (RHS) of (\ref{Tfh}) by the integral $L \int_h^u dg
(g-h) p(g)$ when $L$ is large \cite{Jain:2011d}. Then in the limit $L
\to \infty$, the recursion equation (\ref{PJfns}) reduces to  
\be
{\cal {\bar P}}_{J+1}(f) = (1-\kappa)\int_0^f dh ~\frac{(f-h)~ p(f)}{(1+\kappa
  h)^{\frac{\kappa-1}{\kappa}}}  ~
  {\cal {\bar P}}_J(h)  
\ee
where ${\cal {\bar P}}_J(f) \equiv {\cal P}_J(f, L \to \infty)$. 
A generating function for the 
distribution ${\cal {\bar P}}_J(f)$ can be calculated (see (\ref{Gl})) 
which shows that ${\cal {\bar
    P}}_J(f)$ is finite. Thus from (\ref{QJdef}), it immediately
follows that $Q_J(L) 
\to 0$ as $L \to \infty$ for all $J$. Our numerical results in
Fig.~\ref{behav} for $\kappa=1/2$ show that for $J > 4$, the
distribution $Q_J(L=10^4) < Q_J(L=10^3)$ and for $J <  
4$, $Q_J(L=10^5) < Q_J(L=10^4)$. Thus the distribution $Q_J$
decreases with increasing $L$. 

For $\kappa \geq 1$, the sum in (\ref{Tfh}) can not be replaced by an
integral as the mean of the distribution is infinite. For such
fat-tailed distributions, the sum of $L$ random 
variables is dominated by the largest value ${\tilde f}$ amongst them
\cite{Bouchaud:1990,Sornette:2000}. If at most one fitness exceeds
${\tilde f}$, we  
have $L(1-q({\tilde f})) \sim 1$ or $1+\kappa {\tilde f}=L^\kappa$ for
any $\kappa$. 
Using this result in the recursion equation (\ref{PJfns}) and
changing the variable to $z=(1+\kappa f)/L^\kappa$, we find that for
$\kappa > 1$, 
\bea
{\cal P}_{J+1}(z,L) \propto  \int_{L^{-\kappa}}^{z} dy
(z-y) z^{-\frac{\kappa+1}{\kappa}}
 (1-e^{-y^{-\frac{1}{\kappa}}}){\cal P}_J(y,L)
\eea
where the proportionality constant depends on $\kappa$ and is 
omitted for brevity. Since  the distribution ${\cal
  P}_1(f,L)$ for large $L$ is writeable as 
\be
{\cal P}_1(f,L) \approx \frac{\kappa-1}{L^\kappa} \left( \frac{1+\kappa
  f}{L^{\kappa}} \right) ^{-1/\kappa} =  \frac{\kappa-1}{L^\kappa}
z^{-1/\kappa} 
\ee
it follows that for large $L$, the fitness distribution at the $J$th
step of  adaptive walk is of the following scaling form: 
\be
{\cal P}_J(f,L) \approx \frac{1}{L^{\kappa}} S_J \left( \frac{1+\kappa
  f}{L^{\kappa}} \right) 
\ee
where $S_J(z)$ is a scaling function. 
Using this scaling form in (\ref{QJdef}) and taking the limit $L \to
\infty$, we immediately find that
$Q_J \approx (1/\kappa) \int_0^{\infty} dz 
~S_J(z) ~e^{-z^{-\frac{1}{\kappa}}}$ is finite in
agreement with the numerical results shown in the inset of
Fig.~\ref{behav}.

\section{Walk length cumulants for fitness distributions with finite mean}

For $\kappa < 1$, the probability that the walk terminates {\it at}
the $J$th step is zero or in other words, the
walk goes on indefinitely for infinitely long sequences and hence the
mean number of adaptive steps diverges with $L$. We now show that all the
walk length cumulants increase logarithmically with $L$. 
 
On differentiating (\ref{PJfns}) twice 
with respect to $f$ and writing ${\cal P}_J(f,L)=p(f) P_J(f,L)$, a
straightforward calculation shows that the distribution $P_J(f,L)$ obeys
the following equation \cite{Jain:2011d}:
\be
P^{''}_{J+1}(f,L)=\frac{p(f) (1-q^L(f))}{\int_f^{u} dg (g-f) p(g)} P_J(f,L)~,~J \geq 1
\label{Ppde}
\ee
where prime denotes a $f$-derivative. The boundary conditions are
given by \cite{Jain:2011d}  
\bea 
P_{J}(0,L) = 0 ~,~ P_J'(0,L) = \frac{\delta_{J,1}}{\int_0^u dg ~g ~p(g)}
\label{bc} 
\eea
As (\ref{Ppde}) is non-diagonal in $J$, we work with a generating function  
$G(x,f)=\sum_{J=1}^\infty {  P}_J(f) x^J, x < 1$ 
which obeys the following second order differential equation: 
\be
G''(x,f)= \frac{x p(f) (1-q^L(f))}{\int_f^{u} dg (g-f) p(g)} G(x,f)
\label{Geqn}
\ee

The above differential equation does not appear to be exactly
solvable due to the factor $1-q^L(f)$ on the RHS. As this cumulative
probability decreases from one to zero with increasing $f$, we
consider (\ref{Geqn}) by  approximating 
\begin{numcases}
{1-q^L(f)=1-e^{-{\left(\frac{1+\kappa f}{1+\kappa {\tilde f}}
      \right)}^{-\frac{1}{\kappa}}} \approx} 1 ~,~f < {\tilde f} 
\label{ramp1} \\
r(f)~,~f > {\tilde f} 
\label{ramp2}
\end{numcases}
where $1+\kappa {\tilde f}=L^\kappa$ as found earlier. 
Equation (\ref{Geqn}) has been solved by choosing $r(f)=0$ in 
\cite{Jain:2011d} for $\kappa=-1$ and $\kappa \to 0$. Here we 
show that the leading order behavior of the cumulants does not depend on
the choice of $r(f)$.  
For $f < {\tilde f} $, as a result of (\ref{ramp1}), we have 
\be
G_<''= \frac{x (1-\kappa)}{(1+\kappa f)^2} G_<
\ee
whose solution is of the form $G_<=a_+ (1+\kappa f)^{\alpha_+}+ a_-
(1+\kappa f)^{\alpha_-}$ where  
\be
\alpha_\pm(x)= \frac{1 \pm \sqrt{1+\frac{4 x
      (1-\kappa)}{\kappa^2}}}{2}
\ee
and the constants $a_{\pm}$ can be determined 
using the boundary conditions (\ref{bc}) to finally yield 
\be
G_<= \frac{x (1-\kappa)}{\sqrt{\kappa^2+4 x (1-\kappa)}} \left[(1+\kappa f)^{\alpha_+}- (1+\kappa f)^{\alpha_-} \right]
\label{Gl}
\ee
For $f > {\tilde f}$, using (\ref{ramp2}) in
  (\ref{Geqn}), we get 
\be
G_>''= \frac{x (1-\kappa) r(f)}{(1+\kappa f)^2}
G_>
\label{Glarge}
\ee
whose solution is of the form
\be
G_>= b_1 g_1(x,f)+b_2 g_2(x,f) \label{Gg}
\ee
where the functions $g_1, g_2$ obey (\ref{Glarge}) and $b_1, b_2$ are
constants.  In order to 
compute the walk length cumulants for large $L$, it is 
sufficient to find the $L$ dependence of $b_1, b_2$. This can be done  
by matching the  
solutions $G_<$ and $G_>$ and their first derivative at $f={\tilde
  f}$ and we find 
\bea
b_1(x,L) &=& L^{\kappa \alpha_+} b_{11}(x)+L^{\kappa \alpha_-} b_{12}(x) \\
b_2(x,L) &=& L^{\kappa \alpha_+} b_{21}(x)+L^{\kappa \alpha_-} b_{22}(x) 
\eea
where $b_{ij}(x)$ are independent of $L$.

We now use (\ref{Gl}) and (\ref{Gg}) to write down an expression for the
generating function 
$H(x,L)=\sum_{J=1}^\infty Q_J(L) x^J$ of the walk length
distribution. Using (\ref{ramp1}) in (\ref{QJdef}), we
have 
\bea
H(x,L) &\approx& \int_{\tilde f}^u dh ~(1-r(h)) ~p(h) ~G_>(x,h) \\
& \propto &\int_1^{\frac{1+\kappa u}{L^\kappa}} \frac{dz}{L} ~z^{-\frac{\kappa+1}{\kappa}} ( 1- r(z))
 G_>(x,z) \\
&=& L^{\kappa \alpha_+-1} T_1(x)+
 L^{\kappa \alpha_--1} T_2(x)
\label{Hexpr}
\eea
where the integral 
\be
T_i(x) \propto \int_1^{\frac{1+\kappa u}{
    L^\kappa}} dz ~z^{-\frac{\kappa+1}{\kappa}} ~( 1- r(z))~ (b_{1i}
 g_1(z)+ b_{2i} g_2(z)) \nonumber
\ee
is independent of $L$ which can be seen using the upper
bounds namely $u=-1/\kappa$ for $\kappa < 0$ and infinity for $\kappa
\geq 0$. 
Since $\kappa \alpha_{\pm}-1 < 0$ for any $\kappa$, on taking the
limit $L \to \infty$ in (\ref{Hexpr}), we find that the walk length
distribution vanishes as discussed
earlier. 

\begin{figure}
\begin{center}
\includegraphics[width=0.7 \linewidth,angle=270]{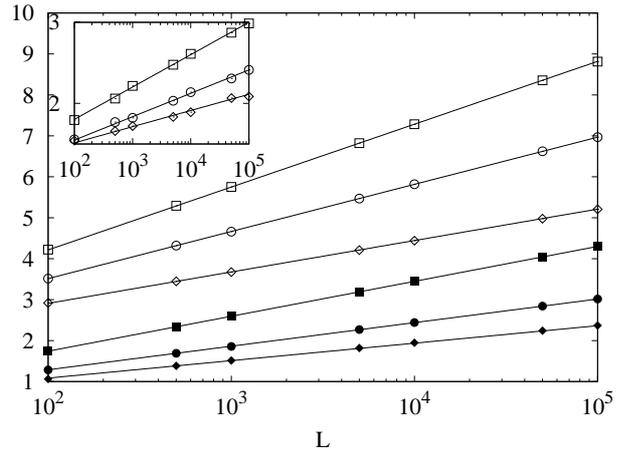}
\caption{Plot of the first three cumulants as a function of sequence
  length $L$ for fitness distributions $p(f)=1$ (squares), $e^{-f}$
  (circles) and 
  $(1+0.5 f)^{-3}$ 
  (diamonds). The main figure
  shows the simulation data for mean $c_1$ (open symbols) and variance
  $c_2$ (filled 
  symbols) and the inset shows the third cumulant $c_3$. The slope of
  the solid lines is given by the 
  analytical results in (\ref{c1})-(\ref{c3}). The 
  numerical data has been averaged over $10^6$ 
  independent realisations of fitnesses and the data for
  $c_2$ has been shifted by a constant for clarity.}     
\label{mom123}
\end{center}
\end{figure}

The fact that $T_i(x)$ is independent of $L$ leads to a
considerable simplification of the problem and allows us to find the
cumulants to leading order in sequence length.  
The $n$th cumulant is defined as \cite{Sornette:2000}
\be
c_n(L) = \frac{d^n \ln H(x,L)}{ds^n}\bigg|_{s=0}
\ee
where $s=\ln x$. As the first term on the RHS of
(\ref{Hexpr}) decays less rapidly than the second term for any
$\kappa$, we have  
$H(x) \approx  L^{\kappa \alpha_+-1} T_1(x)$. Using this, we
immediately obtain the cumulants to leading order in $L$ as 
\be
c_n \approx \frac{\ell}{2} \frac{d^n}{ds^n}
  {\sqrt{\kappa^2+4 e^s (1-\kappa)}}\bigg|_{s=0}~,~n > 0
\ee
where $\ell=\ln L$. Thus we find that all the walk length cumulants
increase logarithmically with $L$. The first three cumulants computed
using the last expression are given by
\bea
c_1 & \approx & \frac{1-\kappa}{2-\kappa} \ell \label{c1} \\
c_2 &\approx & \frac{(1-\kappa) (2-2 \kappa+\kappa^2)}{ (2-\kappa)^3} \ell 
\label{c2}  \\
c_3 &\approx& \frac{(1-\kappa) (4 - 8 \kappa + 6 \kappa^2 - 2 \kappa^3 +
  \kappa^4)}{ (2-\kappa)^5} \ell \label{c3} 
\eea
In the limit $\kappa \to -\infty$, all the above cumulants are equal
to $\ln L$ in agreement with the results for random adaptive walk
\cite{Flyvbjerg:1992}. We also recover the previous results for 
uniformly and exponentially distributed fitnesses
\cite{Jain:2011d}. Equations (\ref{c1}) and (\ref{c2}) also match the 
results of \cite{Neidhart:2011} in which a 
fixed set of mutants during the entire walk is assumed. In contrast,
we have considered a more realistic mutation scheme in which a novel 
set of mutants are available to the population at each adaptive 
step. The above expressions for $c_1$ and $c_2$ have also been seen in
a deterministic model of evolution \cite{Sire:2006} and a relationship
of this model to adaptive walks has been recently elucidated
\cite{Neidhart:2011}. Figure \ref{mom123} shows that  
our expressions (\ref{c1})-(\ref{c3}) agree very well with the
numerical results.

\section{Discussion}

In this article, we studied a biologically realistic model of
adaptation and showed that to leading orders in $L$, the average walk
length is a constant for fitness distributions
with infinite mean but increases logarithmically with the sequence
length otherwise. Our analytical results agree well with the numerical
simulations.  

Our broad theoretical result that the adaptive walks are short (see
Fig.~\ref{mom123}) is consistent with the experiments on microbes
\cite{expt} and fungus 
\cite{Schoustra:2009} in which $2-6$ adaptive substitutions have been
observed. However more detailed experimental studies are needed to
test our predictions. Our result (\ref{c1}) shows that the walk should
last longer in 
 systems with smaller $\kappa$. This may be checked by measuring the
 mean walk length in populations with $\kappa=-1$ \cite{Rokyta:2008},
 $\kappa \to 0$ \cite{kappa0} and $\kappa > 0$ \cite{Neidhart:2011}. 
 To find the dependence of walk length
properties on $L$,  varying the sequence length may not be
experimentally viable but  it should be possible to set up 
experiments along the lines of \cite{Rokyta:2005} and vary the 
initial fitness rank. If the initial ranks are 
of the order $L$, we expect our analysis to hold
\cite{Orr}. Experimental data for the walk length distribution 
showing insensitivity to the  
initial rank would then imply an underlying fat-tailed fitness
distribution with infinite mean.

\section{Acknowledgement} The author thanks J. Krug for useful
comments on the manuscript. 

\end{document}